\documentclass[11pt]{article}

\usepackage{amsmath,amsfonts,amsthm,epsf,stackrel}

\usepackage{graphicx,subfigure}
\usepackage{setspace}
\makeatletter
\usepackage{enumerate}
\usepackage{float}
\def\baselinestretch{1.5}

\usepackage[margin=2cm]{geometry}
\usepackage[xcolor=dvipsnames]{xcolor}

\usepackage{tcolorbox}

\usepackage{dsfont}
\usepackage{multirow}

\usepackage{booktabs} 

\usepackage{hyperref}
\hypersetup{
	colorlinks=true,
	linkcolor=blue,
	filecolor=magenta,      
	urlcolor=cyan,
	pdftitle={Overleaf Example},
	pdfpagemode=FullScreen,
}

\newcommand\blfootnote[1]{%
	\begingroup
	\renewcommand\thefootnote{}\footnote{#1}%
	\addtocounter{footnote}{-1}%
	\endgroup
}

\begin{document}
\title{Surgery duration prediction using multi-task feature selection}

\author{David Azriel, Yosef Rinott, Orna Tal, Benyamine Abbou, and Nadav Rappoport}

\maketitle
\blfootnote{
	D.A. - Faculty of Data and Decision Sciences, The Technion - Israel Institute of Technology, Israel (e-mail: davidazr@technion.ac.il)\\
	Y.R - Department of Statistics and Federmann Center for the Study of Rationality, The Hebrew University of Jerusalem, Israel (e-mail: yosef.rinott@mail.huji.ac.il).\\
	B.A -  Medical Management Department, Hillel-Yaffe Medical Center, Israel (e-mail: benyaminea@hymc.gov.il).\\
	O.T -  Medical Management Department, Shamir Medical Center (Assaf Harofeh), Israel (e-mail: ornatal@shamir.gov.il).\\
	N.R -  Department of Software and Information Systems Engineering, Ben-Gurion University of the Negev, Israel (e-mail: nadavrap@bgu.ac.il).\\
	Y. R. was supported in part by a grant from the Center for Interdisciplinary Data Science Research at the Hebrew University (CIDR).}
\begin{abstract}
Efficient optimization of operating room (OR) activity poses a significant challenge for hospital managers due to the complex and risky nature of the environment. The traditional ``one size fits all" approach to OR scheduling is no longer practical, and personalized medicine is required to meet the diverse needs of patients, care providers, medical procedures, and system constraints within limited resources. This paper aims to introduce a scientific and practical tool for predicting surgery durations and improving OR performance for maximum benefit to patients and the hospital.
		Previous works used machine-learning models for surgery duration prediction based on preoperative data. The models consider covariates known to the medical staff at the time of scheduling the surgery. 
		{ Given a large number of covariates, model selection becomes crucial, and the number of covariates used for prediction depends on the available sample size.}
		 Our proposed approach utilizes multi-task regression to select a common subset of predicting covariates for all tasks with the same sample size while allowing the model's coefficients to vary between them. A regression task can refer to a single surgeon or operation type or the interaction between them. By considering these diverse factors, our method provides an overall more accurate estimation of the surgery durations, and the selected covariates that enter the model may help to identify the resources required for a specific surgery.
		We found that when the regression tasks were surgeon-based or based on the pair of operation type and surgeon, our suggested approach outperformed the compared baseline suggested in a previous study. However, our approach failed to reach the baseline for an operation-type-based task. 
		By accurately estimating surgery durations, hospital managers can provide care to a greater number of patients, optimize resource allocation and utilization, and reduce waste. This research contributes to the advancement of personalized medicine and provides a valuable tool for improving operational efficiency in the dynamic world of medicine.
\end{abstract}


\section{Introduction}
\label{sec:introduction}
	Hospital managers frequently cope with the challenge of optimizing operating room (OR) activity. This is a complex undertaking involving many factors in a risky environment. From the managers' perspective, OR activity is not only costly but also demanding on a professional level, necessitating the coordination of multidisciplinary staff and highly skilled surgeons. Still, it encompasses high profitability as well as the opportunity to achieve professional excellence. Therefore, managers desire expansion of the variety of procedures as well as innovative activity, and tend to encourage physicians to increase their performance.
	
	The complexity of OR activity faces two major challenges: The first is related to the diversity of the surgeons in terms of their skills and experience in different procedures. The second is related to the environmental features, patients' characteristics, diversity of staff members, and multiple infrastructure components, all of which lead to many potential scenarios. These features increase the uncertainty of achieving maximal efficiency.
	
	Structured plans based on ``one size fits all" are no longer effective practical solutions. Medical managers look for ``personalized medicine" to meet the patient's needs and preferences, the care provider’s characteristics, the procedure's requirements, and the system's constraints within scarce resources. The advantages of personalized medicine are twofold. First, it enables better resource allocation and utilization and reduces waste, which in turn enables the provision of care to additional patients. However, in the current practice, medical managers lack the ability to consider all relevant elements within the dynamic world of medicine and may fall short of an accurate estimation of the resources required for a specific surgery. 
	The aim of this paper is to introduce a scientific yet practical tool to predict the duration of surgeries in order to improve OR performance and maximize the benefit for both patients and hospitals.

	OR scheduling is traditionally based on a staff member’s estimation of operation duration based on personal or institutional previous experience, yet the accuracy of such estimations is quite low \cite{laskin2013}. Solutions to maximize OR performance have been discussed (\cite{di2014assessment,hanset2010using,pham2008surgical,lin2021solving,martinez2019process,antognini2015many})  and different machine learning (ML)-based approaches have been developed to pursue an advanced methodology for managing OR utilization (\cite{bartek2019improving,abbou2022optimizing}).
	The ML-based approaches require training data and are based on { the use of statistical models}. The advantage of ML-based approaches vs. rule-based approaches is that the former are data-driven as opposed to knowledge-driven. In addition, ML-based approaches can take into account complex relationships between data points. The computational models are trained on large datasets representing past experience and are driven by algorithms to accurately predict unknown labels or future events. 
	

In a previous paper (\cite{abbou2022optimizing}) we developed an ML method for predicting surgery duration using preoperative data of the planned operations based on the eXtreme Gradient Boosting (XGB) model. We trained and evaluated our method using data from a large general public hospital in Israel. The method uses only covariates that are known to the medical staff at the time of scheduling the surgery, which is the day before the surgery starts, and fits a single large model for the entire dataset. Here we study a different approach that considers a multi-task regression prediction and is based on the work of \cite{azriel2021optimal}. The idea is to choose a common subset of predicting covariates for all tasks having the same sample size but allowing the model's coefficients to vary between tasks.
	A regression task can refer to a single surgeon, to operation type, or to the interaction between them. Using a model selection procedure that accounts for the number of observations is natural as a large sample size allows a larger subset of covariates to be used for prediction. In order to obtain a good prediction, a few covariates should be used with a small sample size, and a more complex model with more covariates can be fitted when the sample size is large, {provided, of course, that a suitable method is employed to verify that the additional variables improve the prediction}.

{In general, the aim of feature selection is to remove noisy relatively
non-informative features in order to avoid overfitting of the statistical
model, and guarantee generalizability. In order to place our data analysis
in the machine learning literature we briefly discuss three feature
selection approaches: 1) filters; 2) wrappers; and 3) embedded methods.  We
limit the number of references provided here; many further  references
related to this brief review can be found in \cite{bolon2013} and \cite{pudjihartono2022}.}

{Filter methods consist of feature selection based on the correlation, or
other measures and tests of association between each feature to be selected
or removed and the variable to be predicted. This selection method is
independent of the prediction model being used. Our multi-task approach
starts with preprocessing filtering aimed at identifying the key features of
a given prediction problem. We use univariate feature filtering based on
correlation, with the purpose of removing variables that are clearly
irrelevant to our prediction purpose, thus reducing the complexity of the
multivariate model selection method that follows this first stage.}

{The second stage of our multi-task approach can be seen as a special case
of the wrapper methods, in which the performance of subsets of features is
compared relative to a given prediction model, and the best subset is
chosen. In the multi-task approach, we estimate the performance of each
subset of the features that passed the first-stage filter using a linear
regression model based on the features contained in the subset.}
{We then choose the best-performing subset. This is a multivariate approach
as we compare whole subsets of features rather than individual ones. In the
presence of a large number of features, this approach is computationally
demanding; in our case, we were left with a manageable number of features
after the first-stage filtering.}

{The XGB approach of \cite{abbou2022optimizing}, which was mentioned above, is an example of the embedded methods, where the tree (and more generally, the model) is
constructed along with the feature selection. In general, these methods are
computationally less demanding than wrapper methods. We found that for most
of the prediction undertakings, the multi-task approach performs better than
the XGB model.}

\section{Methods}
	
	\subsection{Data} \label{sec:data}
The dataset is extracted from the electronic health records of a large (891-bed) public hospital (Shamir Medical Center). The data were accessed through the Kineret platform of the Israeli Ministry of Health. The Kineret platform made de-identified data from the Israeli medical centers accessible for research in the format of the Observational Medical Outcomes Partnership (OMOP) Common Data Model (CDM).  The dataset contains all surgeries that occurred from December 2009 to May 2020. In this work, we focus on data from two general surgery departments comprising a total of 23,183 surgeries, 146 lead surgeons, and 2,381 types of operations. As in \cite{abbou2022optimizing}, the data was split into training and test sets such that surgeries from 2018 and on were used only for testing. Our focus in this paper is on models that allow different coefficients for each lead surgeon or operation type. Therefore, we consider a subset of the data which contains 32 lead surgeons with more than 100 surgeries and 123 types of operations with more than 15 observations in the training data. This subset contains 13,359 and 3,028 observations in the training and test data, respectively. 

In this study, we aim to scrutinize models that predict surgery duration. Since the distribution of surgery duration has a long right tail, we consider the natural log of durations instead of the durations. When the difference between log durations (multiplied by 100) is small, the difference is approximately the difference in terms of percentage; see Chapter 2.4 in \cite{hansen2022econometrics}. Thus, in the present study, the prediction errors are interpreted as the difference in percentage between the predicted duration of the task and its actual duration.  

\begin{figure*}[!t]
	\centerline{\includegraphics[width=\textwidth]{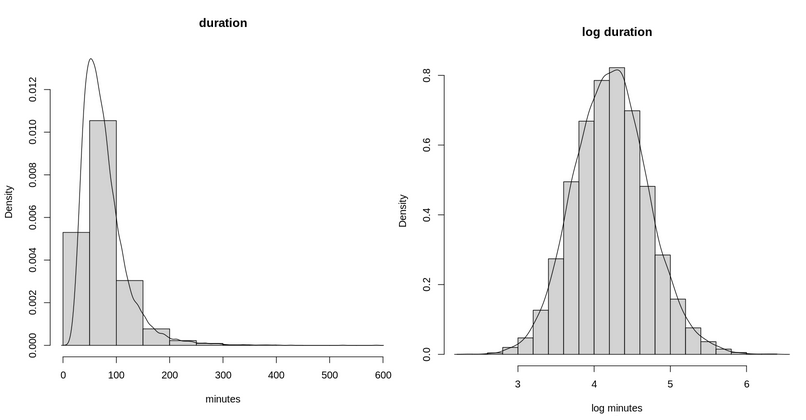}}
    \caption{\footnotesize Histogram and density estimates of the duration and log duration in the training data.}
    \label{fig:hist}
\end{figure*}

Table \ref{tab:1} provides summary statistics for the training and test data, and Figure \ref{fig:hist} plots the histogram and density estimates in the training data of the surgerys' durations. It is demonstrated that under the log transformation, the distribution of the observations is close to a normal distribution. 
The summary statistics of the training and test data are generally similar, except for the surgeon's experience, which is higher in the test data as the experience increases with time; also, the duration is slightly higher in the test data. 


 \begin{table}
 	\begin{center}
    \caption{  Summary statistics of the training and testing data. $N$: Number of surgeries. IQR: Interquartile Range.}
		\label{tab:1}
  \setlength\tabcolsep{2pt}
		\begin{tabular}{|l|c|c|}
			\hline
			{} & Training ($N$=13,359) &  Test ($N$=3,028) \\
			\hline
			\hline
			Demographics: & & \\
			Age (median, IQR) & 53 (37--66) & 54 (39--68) \\
			Female (\%) &  47.4 \% & 51.5\% \\
			\hline
			Preoperative: & &\\
			\# Drugs (median, IQR) & 6 (3--13) & 5 (3--11) \\
			\# Diagnosis (median, IQR) & 7 (3--13) & 8 (3--16) \\
			\hline
			Surgeon's experience: & &\\
			\# Previous surgeries (median, IQR) & 437 (198--817) & 1382 (573--1676)\\
			Previous hours (median, IQR) & 559.5 (260.4--1134.2) & 1990 (891--2495)\\
			\hline 
			Surgery: & &\\
			Duration in minutes (median, IQR) & 68.1 (49.1--94.1) & 75.0 (56.4,99.4)\\
			Log duration (median, IQR) & 4.22 (3.89--4.54) & 4.32 (4.03--4.60)\\
			\hline
		\end{tabular}	
\end{center}
	\end{table}

\begin{figure*}[!t]
	\centerline{\includegraphics[width=\textwidth]{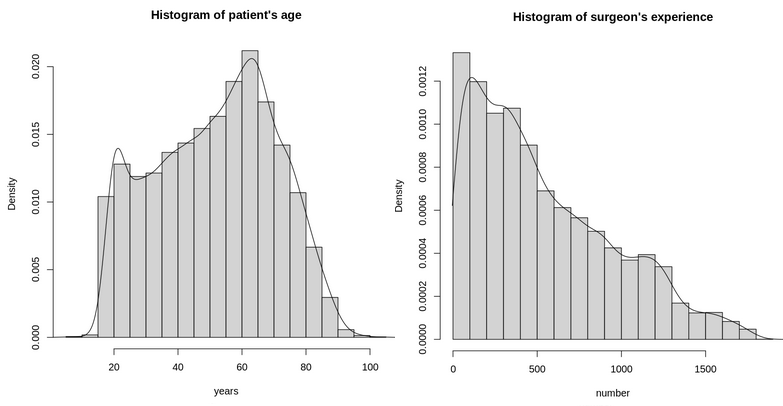}}
	\caption{\footnotesize Histogram of the variables ``patient's age" and ``surgeon’s experience" (total number of past surgeries) in the training data.}
    \label{fig:hist_cov}
\end{figure*}

In the dataset, there are 972 covariates that could be used to predict surgery duration.
{A description of the covariates is given in \cite{abbou2022optimizing} and a full list of the covariates can be found in the supplementary table of the latter paper.
	Most of the covariates have little or no prediction power.} 
As a preprocessing step, only a small number of them were selected. First, a linear regression model was considered with the two categorical variables, lead surgeon and operation type. Then, the six most correlated variables with the residuals of the model were considered. These variables were (ordered by the magnitude of the correlation): patient's age, number of anesthesiologists, hypertension, Ot.compl.bir {(a CCS category of diagnosis codes for Other complications when the patient is a mother during puerperium)}, diabetes mellitus (without complications), and surgeon’s experience (total number of past surgeries). 
Histograms of the variables ``patient age" and ``lead surgeon’s total number of past surgeries" in the training data are plotted in Figure \ref{fig:hist_cov}  and the distribution of the other variables in the training data are presented in Table \ref{tab:2}.

Another variable that is highly correlated with the residuals is the number of nurses. However, it turns out that when adding it to the model, the predictions are highly biased in the test set. This may be due to policy changes in the years of the test data (2018-2020) that are aimed at reducing the number of nurses assigned to uncomplicated operations due to a shortage of qualified nurses. {Thus, our models are based on external facts such as hospital policy and could perform poorly if changes occur. However, it is clear that prediction must always be based on the past, and assume some stationary behavior. }

\begin{table*}
     \caption{ The distributions of the variables in the training data; frequency (in percentage) is in parentheses.}
	\label{tab:2}
 \centering
    \begin{tabular}{|c||c|c|c|c|c|}
		\hline
		Variable $\big \backslash$ Value & 0 & 1 & 2 & 3 & 4 \\ \hline \hline
		\# Anesthesiologists & 5112 (38.3\%) & 5679 (42.5\%) & 2455 (18.4\%) & 110 (0.8\%) & 3 (0.02\%)\\
		\hline
		Diabetes Mellitus & 11579 (86.8 \%) & 1762 (13.2 \%) & & &  \\
		\hline
		Hypertension & 9451 (70.7\%) & 3908 (29.3 \%) & & &\\
		\hline
		Ot.compl.bir & 12365 (92.6 \%) & 994 (7.4 \%) & & &\\
		\hline
	\end{tabular}
\end{table*}

\subsection{Methodology}

Our predictions are computed by a multi-task regression model.
Since variable selection for prediction is challenging as it requires a large sample size, the approach of \cite{azriel2021optimal} is to choose, on the basis of the training dataset, a common set of covariates for each sample size. On the other hand, estimation for each prediction task is based on a relatively small subset of the dataset; for example, for prediction of future surgery durations of each particular surgeon, only data pertaining to this surgeon is used in estimating the prediction model.  

More specifically, \cite{azriel2021optimal} consider a dataset consisting of $J$ regression tasks. For each regression task $j \in \{1,\ldots,J\}$, let $R_j(n,p)$ denote the prediction error of the linear model $p$ when the sample size is $n$. The quantity of interest is $R(n,p)=\frac{1}{J} \sum_{j=1}^J R_j(n,p)$, which is the average prediction error of model $p$ over the $J$ tasks if they had a common sample size $n$. Azriel and Rinott define a statistic $C^{(p)}(n)$ (see Eq. (3.4) in that paper), which is a generalization of Mallows's $C_p$ and it is approximately unbiased and consistent for the average prediction error under certain conditions. {The $C^{(p)}(n)$ statistic is composed from the residual sum of squares plus a ``penalty term", which depends in particular on the number of predictors as well as the sample size. This penalty term is needed to avoid bias in the estimation of $R(n,p)$ and also to prevent overfitting. }
For each sample-size $n$, one can estimate the best model for $n$ by $\arg \min_p C^{(p)}(n)$; for the asymptotic properties of this estimator see Sections 3.2 and 3.3 in \cite{azriel2021optimal}.

Thus, the training data is divided into subsets, where each subset is considered a different regression task. For each task, a separate linear regression model is estimated, but the variable selection is common across different tasks, in the sense that two tasks with the same sample size use the same set of covariates (but different coefficients). 

We consider two ways of making data subsets and corresponding regressions – either estimating regression coefficients separately for each lead surgeon or for each combination of the lead surgeon and the operation type – and compare them to a single global regression model. We also compare these models' predictions to predictions given by a complex model based on the XGB model as described in \cite{abbou2022optimizing}. A third way of making subsets was considered, namely, by the operation type, but this led to higher prediction errors and will not be described here.

\noindent{\bf Global regression model:} A single linear regression model was considered using the whole training set, where the response variable was the log duration, and the covariates were: lead surgeon and operation type, which are categorical variables, and the six covariates mentioned above. The parameters of the model were estimated using least squares.

\noindent{\bf Surgeon-based models:} For every lead surgeon a different regression model was considered. As a first step, the response variable for the $i$-th observation was defined to be $Y_i-\bar{Y}_i$, where ${Y}_i$ is the log duration, and $\bar{Y}_i$ is the average of the log durations of the operation type of the $i$-th observation{, where the average is over all the observations in the training set of this operation type}. For each surgeon, a separate regression model was considered, where the covariates are: operation type (if there are more than 10 observations from the same type) and the six covariates mentioned above. The variable selection procedure was based on the model selection procedure of \cite{azriel2021optimal}.

\noindent{\bf  Interaction-based models:} Here, a task is defined by a pair consisting of a certain lead surgeon and an operation type. {In these models the response variable is the log duration, i.e., $Y_i$, and not the difference. This is because in the interaction models, each task is associated with one operation type and therefore the operation type is implicitly included in the intercept of the model.}
There are 193 such tasks each having at least 15 observations in the training data. Out of the 13,359 observations of the training data, the total number of observations in these tasks is 9,129. That is, about 4,000 observations are excluded from the analysis under these models.  

\subsection{Evaluation metrics}

The prediction square error was defined to be the square of the difference between the log of the predicted duration and the log of the actual duration. The mean prediction error for each sample size was estimated by the $C^{(p)}$ statistic that is defined in Eq. (3.4) in \cite{azriel2021optimal}. {For each task, the mean squared prediction error was evaluated in the training data by 10-fold cross-validation repeated 500 times. For each repetition, 90\% of the data was used for the estimation of the parameters and 10\% for estimating the prediction error. An average over the 500 repetitions was computed. For the test data, the actual prediction error was used.} Some tasks appear only in the training data and not in the test data, and for such tasks, the actual prediction error is missing. We report the root mean squared prediction error of the log of the durations, denoted by RMSE, which represents an approximation to the percent deviations from the actual durations, as explained above.

\section{Results}

\subsection{Global regression model} We estimated the coefficients of the model based on the training data. The fraction of the variance explained by the model ($R^2$) is 0.62, indicating a relatively large variance of the response explained by the model. However, most of the explained variance comes from the (categorical) variable ``operation type". The $R^2$ statistic of a linear regression model with this single variable is 0.55.
The RMSE in the training data, based on a 10-fold cross-validation, is 30.1\% and in the test data, the corresponding number is 30.0\%.

\subsection{Surgeon-based models}  There there are 32 tasks corresponding to 32 lead surgeons. As usual in the present methodology, the variable selection criterion is sample size-dependent. For all tasks, the categorical variable ``operation type" (for types with more than 10 observations) is included and among the six covariates, the variable selection is given in Table \ref{tab:3}.

\begin{table}
	\centering
    \caption{The covariate selection procedure as a function of the sample size of the task ($n$).}
    \label{tab:3}
	\begin{tabular}{|c||c|}
		\hline
		Task sample size & Covariates \\ \hline \hline
		$52 \le n \le 192$ &  \# Anesthesiologists, patient age, surgeon’s experience\\
		\hline
		$193 \le n \le 1052$  &  \# Anesthesiologists, patient age, surgeon’s experience, Ot.compl.bir\\
		\hline
	\end{tabular}
\end{table}

\begin{figure*}[!t]
	\centerline{\includegraphics[width=\textwidth]{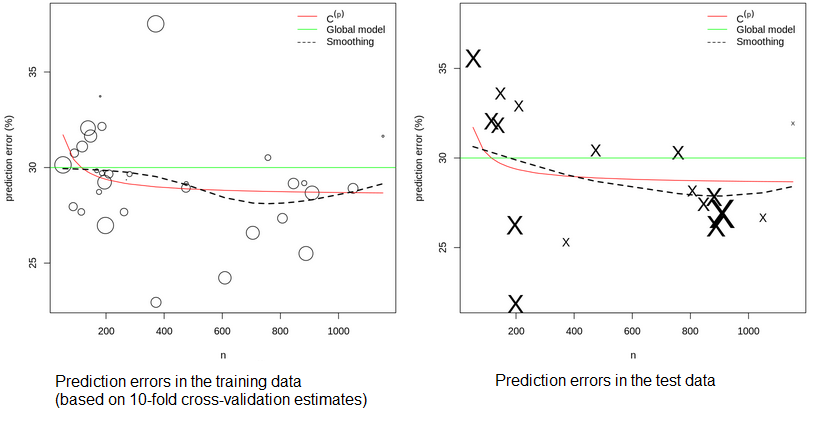}}
	\caption{\footnotesize The prediction errors in the training data (left panel) and in the test data (right panel) as a function of the sample size for the surgeon-based models. Each mark in the plot represents prediction errors of surgeries performed by the same surgeon. For the training data, the prediction errors are based on the 10-fold cross-validation. The size of the circles and the crosses is proportional to the mean of the log duration. The red line is the $C^{(p)}$ estimate of the mean prediction error. The green line is the prediction error of the global regression model and the black dashed line is a kernel smoothing of the prediction errors.}
    \label{fig:surg}
\end{figure*}

	Figure \ref{fig:surg} plots the prediction errors in the training and test data as a function of the sample size. Recall that in the training data, the prediction error is estimated by a 10-fold cross-validation. Only 17 tasks out of 32 appear in the test data. Also plotted is the $C^{(p)}$ estimate of the average prediction error of \cite{azriel2021optimal} and a kernel smoothing of the prediction errors. As a benchmark, the prediction error of 30.0\% of the global regression model is also given. 

It is demonstrated that the different estimates of the prediction errors as well as the actual prediction errors are generally in agreement. In particular, the black dashed line, which is a kernel smoothing of the prediction errors, is close to the red line, which is the  $C^{(p)}$ estimate of the average prediction error. One can also observe that on average the surgeon-based models yield slightly smaller prediction errors than the global regression model, and this is especially true for tasks with a large sample size. Interestingly, there seems to be no clear connection between the mean duration of the task and the prediction error.

	\subsection{Interaction-based models}
Recall that in these models, a task is defined by a  lead surgeon and an operation type that together have 15 observations or more in the training data; there are 193 such tasks. Here, since for each task the operation type and lead surgeon are the same, the only covariates of the model are the six covariates mentioned above. The variable selection procedure as a function of the sample size is given in Table \ref{tab:4}.

\begin{table}
	\begin{center}
    \caption{The covariate selection procedure as a function of the sample size of the task (n).}
	\label{tab:4}
	\begin{tabular}{|c||c|}
		\hline
		Task sample size & Covariates \\ \hline \hline
		$15 \le n \le 27$ & surgeon’s experience\\
		\hline
		$28 \le n \le 63$  & surgeon’s experience, patient age\\
		\hline
		$64 \le n \le 445$  & surgeon’s experience, patient age, \# Anesthesiologists\\
		\hline
	\end{tabular}
\end{center}
\end{table}

\begin{figure*}[!t]
    \centerline{\includegraphics[width=\textwidth]{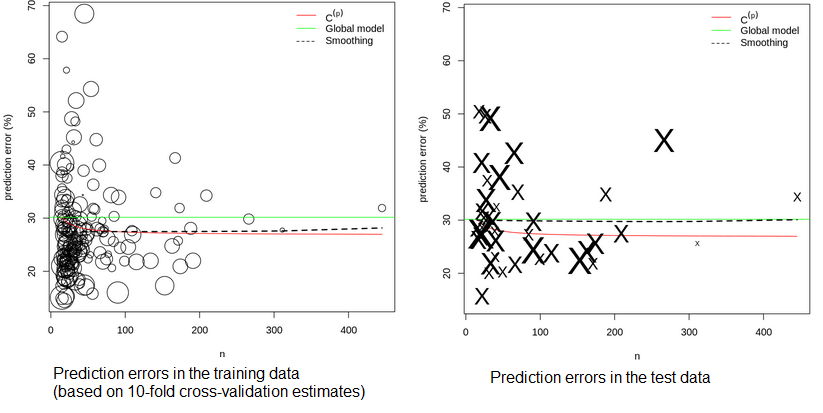}}
	\caption{\footnotesize The prediction errors in the training data (left panel) and in the test data (right panel) as a function of the sample size for the interaction-based models. Each mark in the plot represents prediction errors of surgeries performed by the same surgeon. For the training data, the prediction errors are based on the 10-fold cross-validation. The size of the circles and the crosses is proportional to the mean of the log duration. The red line is the $C^{(p)}$ estimate of the mean prediction error. The green line is the prediction error of the global regression model and the black dashed line is a kernel smoothing of the prediction errors.}
    \label{fig:inter}
\end{figure*}
Figure \ref{fig:inter} parallels Figure \ref{fig:surg} for the interaction-based models. In the test data, we considered only tasks with 10 or more observations and there are 42 out of 193 such tasks. Notice that for the small samples (small $n$), some of the errors were very large. Similar to the surgeon-based models, for the training data the $C^{(p)}$ estimate of the mean prediction error and the kernel smoothing of the prediction errors generally agree. The average prediction error is less than the 30\% of the global regression model. However, in the test data, the average prediction error was 30.1\%,  and is slightly higher than the $C^{(p)}$ and cross-validation estimates. One possible explanation is that the $C^{(p)}$ and cross-validation estimates require a large sample size, whereas in these models the sample sizes are relatively small, and therefore the estimates may be highly variable or biased. 

\subsection{{Comparisons to other models and methods}}

{We compared the prediction errors of the surgeon-based models to the eXtreme Gradient Boosting (XGB) model previously used in \cite{abbou2022optimizing} for this dataset as well as to three other feature selection methods: LASSO (\cite{tibshirani1996}), mutual information (\cite{battiti1994}), abbreviated MI, and forward selection, abbreviated FS. Notice that the XGB model was trained on the entire dataset of all hospital departments and here we evaluate its performance only for the two general surgeries departments we considered in this work.  
	For LASSO, MI, and FS we used the two categorical variables lead surgeon and operation type, which amounts to 153 dummy variables and all other numerical variables in the dataset with no missing values. The total number of covariates is 495. We computed the LASSO coefficients using the glmnet package in R, which also calculates the the tuning parameter by cross-validation. For MI and FS we selected 200 variables out of 495 and estimated the parameters of the global linear model with respect to these covariates. 
	The same split of the training and test sets as described in Section \ref{sec:data} was considered.} 

\begin{figure*}[t!]
	\centering
	\includegraphics[width=1\textwidth]{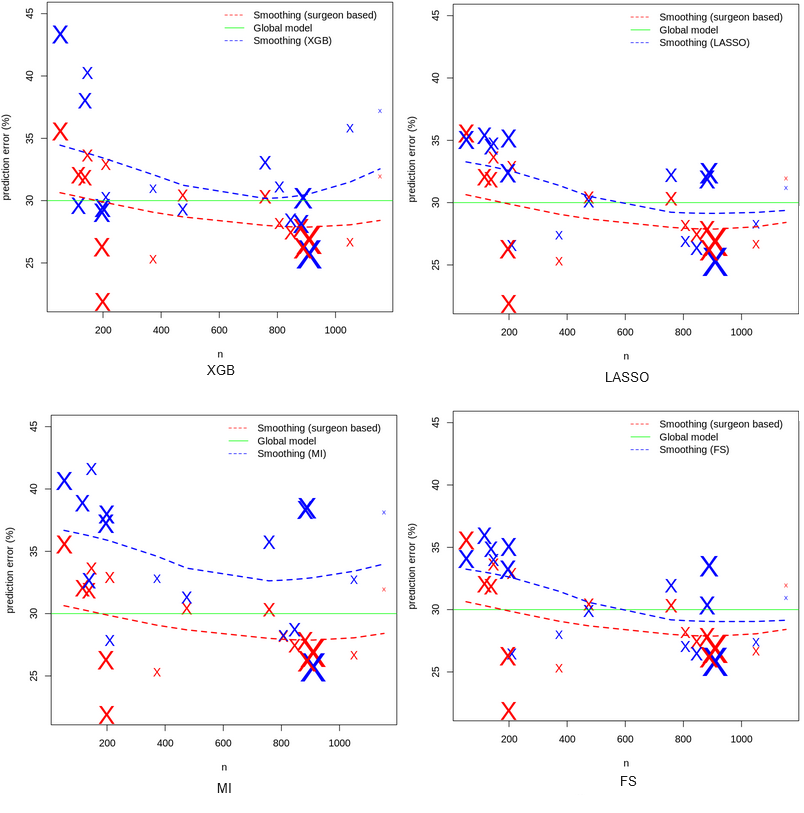}
	\caption{\footnotesize {The prediction errors as a function of the sample size in the test data for the surgeon-based models (red) compared to the XGB model, LASSO, MI and FS (blue). The size of the crosses is proportional to the mean of the log duration. The green line in the prediction error of the global regression model and the red (respectively, blue) dashed line is a kernel smoothing of the prediction errors for the surgeon-based models (respectively, XGB, LASSO, MI and FS).}}\label{fig:comp}
\end{figure*}

{Figure \ref{fig:comp} plots the prediction errors in the test data of the surgeon-based models (red) compared to the XGB, LASSO, MI and FS (blue). The prediction errors in the test set of the surgeon-based models, XGB, LASSO, MI and FS are 29.4, 32.4, 30.1, 33.5 and 30.0, respectively. Thus, the surgeon-based models have smaller prediction error than the other methods we compared and the relative improvement  varies from 2\% to 10\%. It is interesting to note that the smoothing line of the surgeon-based models in Figure 5 is consistently below the other methods for all sample sizes. Another important
 	 advantage of the surgeon-based models is that they use much smaller numbers of covariates allowing for simpler models, better interpretability, and possibly reducing the work required by recording and using more variables. }


\section{Discussion}
Looking closer at the prediction errors we observe that there are two tasks with a relatively high prediction error; one of these tasks corresponds to a particular surgeon, and one to the interaction of a particular surgeon and operation type. 

The first task with high prediction error appears in the surgeon-based models, where each task corresponds to a surgeon. Looking at the left panel of Figure \ref{fig:surg} one can find a task (a surgeon) with a prediction error of 37.5\% in the training data and sample size of 371 operations. This prediction error is extreme for the surgeon-based models. There are no observations for this surgeon in the test data. Focusing on the surgeon of this task, we observe that his/her surgeries are longer compared to those of the other surgeons (second largest) and the average age of his/her patients is high (fourth largest). Considering the other covariates in the data, we see that the average number of surgeons participating in his/her surgeries (excluding the lead surgeon) is the highest, and also his/her surgeries have the highest number of different diagnoses on average. These findings indicate that the surgeries of this lead surgeon were highly complicated or utilized a new technique or device and therefore were prone to complications,  resulting in longer performance duration on average.
{Yet, academic (teaching) hospitals take into account and even encourage such events as part of the strategy to implement innovation or gain trainees' experience.}
Thus, for this surgeon there seems to be a combination of circumstances that causes high prediction errors: his/her surgeries were carried out during a relatively short period, the patients had a variety of medical conditions and many surgeons participated in those operations.

The second task with a relatively high prediction error can be found in the interaction-based models, where each task corresponds to a combination of a lead surgeon and an operation type. This task has a sample size of 266. The prediction error in the training data was 29.6\%, which is close to the expected error for a task with this sample size, but in the testing data the prediction error was 45.0\%, which is quite high; see the right panel of Figure \ref{fig:inter}. Looking closer at the errors, one observes that most of them arise from bias. The average actual duration was 106.7 minutes but the average predicted duration was 72.1 minutes. The reason for this bias is that the test data for this task is quite different from the training data, in which the average duration was 60.9 minutes. {Looking at the covariates for this task in the training data versus the test data we observe significant differences: the patients are older in the test set (average of 61 years versus 45), their drug count is higher (average of 11.9 versus 6.6) and the number of anesthesiologists is higher (average of 1.12 versus 0.76). This implies that the surgeries in the test set, while being of the same type and carried out by the same surgeon, are more complicated and hence their length is harder to predict. }
It seems that no statistical method can prevent prediction errors due to such differences between the training and the test data.  

{The multi-task approach allows one to identify tasks with high prediction errors like the surgeon-based task having a prediction error of 37.5\% mentioned above. For such tasks, one can consider a further division of the task, or use a different prediction method, or just provide a warning that for this task the prediction error is high. This is another advantage of the multi-task approach since the prediction error can be estimated for each task separately rather than providing one prediction error as in the global models. }
 

{Considering the six covariates that were mentioned in Section \ref{sec:data} we see that under the global regression model, all six are statistically significantly different from zero. For \# Anesthesiologists and patient age the value of the t-statistic is larger than 10, for surgeon's experience and Ot.compl.bir it is between 5 and 8 and for Diabetes Mellitus and Hypertension it is about 3. Correspondingly, we see that for the surgeon-based models the covariates \# Anesthesiologists, patient age and surgeon's experience are selected for tasks with sample size smaller than 192 and Ot.compl.bir is added for larger tasks (see Table \ref{tab:3}). For the interaction-based models, the order by which the covariates are selected is surgeon's experience, patient age and \# Anesthesiologists (see Table \ref{tab:4}). }

{In the multi-task models, i.e, the surgeon-based and interaction-based, it is of interest to compare our model selection procedure to a model that selects all six covariates and uses different coefficients in each task. In the surgeon-based models where the sample size of the tasks is relatively large (it varies from 
52 to 1052), the difference between the predictions of the model with six covariates and the suggested one is quite small. The relative prediction improvement of our model, as estimated by cross-validation estimates is smaller than 0.5\%. By the $C^{(p)}$ statistic, one can evaluate the relative improvement for different sample sizes and it turns out that for sample sizes of around 50 the relative improvement is about 2\%. In the interaction models, the sample size of the tasks is smaller (about 90\% of the tasks have fewer than 100 observations) and the relative improvement is more significant; it is about 7\% for the entire training set and about 10\% for tasks with sample sizes of 15.}

	We now make a few general comments comparing our method to other available options. A common approach in ML methods is to train a single model on the entire dataset, under the assumption that a larger sample size will lead to lower prediction error. Regularization methods are used to reduce the generalization error of models. Here, we used a different approach described by \cite{azriel2021optimal} where different model parameters were estimated for different tasks, but the covariate selection is such that all tasks having the same sample size were modeled with the same set of covariates.  We showed that when the tasks correspond to the lead surgeon, the models that use a handful of features perform better than a global linear model and even better than a complex XGB model that is trained using hundreds of features. The model achieved an average prediction error of 28.9\% in the training data (based on 10-fold cross-validation estimates) and 29.2\% in the test data. However, when the tasks correspond to both the operation type and the surgeon, the prediction error in the test data was about the same as in the global linear model. For these models the average prediction error was 27.8\% in the training data (based on 10-fold cross-validation estimates) and 30.1\% in the test data.

	In general, ML and statistical methods can be used in healthcare systems to analyze large data and to help managers understand the operation activity and improve it. The use of such methods in healthcare is evolving and researchers as well as managers still learn how to use and implement those methods. In the present work, we considered a feature selection procedure that was suggested for multi-task regression and studied its prediction performance. Further research is required to fully understand the potential of this method.

\section*{Conflict of interest} None of the authors have a conflict of interest to disclose.

\section*{Institutional Review Board Statement}
The study was conducted according to the guidelines of the Declaration of Helsinki and approved by the Institutional Review Boards of Shamir Medical Center (protocol code 0308-19-ASF on 3 December 2019).

\bibliographystyle{chicago}
\bibliography{main}

\end{document}